\documentclass{iopart}
\usepackage{iopams}
\usepackage{dsfont}
\usepackage{amsthm}
\usepackage{graphicx}
\usepackage{bbm}
\usepackage{cite}

\newcommand{\bra}[1]{\langle #1|}
\newcommand{\ket}[1]{|#1\rangle}
\newcommand{\braket}[2]{\langle #1|#2\rangle}
\newcommand{\eval}[1]{\langle #1\rangle}

\newcommand{\dbra}[1]{\langle #1||}
\newcommand{\dket}[1]{||#1\rangle}
\newcommand{\dbraket}[2]{\langle #1||#2\rangle}

\newcommand{\la}{\langle}
\newcommand{\ra}{\rangle}

\newcommand{\rf}[1]{(\ref{#1})}

\renewcommand{\Re}{{\rm Re}}

\begin{document}
\title[Mean-field approximation for a Bose-Hubbard dimer with complex interactions]{Mean-field approximation for a Bose-Hubbard dimer with complex interaction strength}

\author{Eva-Maria Graefe${}^1$ and Chiara Liverani${}^{1,2}$}
\address{${}^1$ Department of Mathematics, Imperial College London, London, SW7 2AZ, United Kingdom\\
${}^2$ Department of Physics, Imperial College London, London, SW7 2AZ, United Kingdom}
\begin{abstract}
In the limit of large particle numbers and low densities systems of cold atoms can be effectively described as macroscopic single particle systems in a mean-field approximation. In the case of a Bose-Hubbard system, modelling bosons on a discrete lattice with on-site interactions, this yields a discrete nonlinear Schr\"odinger equation of Gross-Pitaevskii type. It has been recently shown that the correspondence between the Gross-Pitaevskii equation and the Bose-Hubbard system breaks down for complex extensions. In particular, for a Bose-Hubbard dimer with complex on-site energy the mean-field approximation yields a generalised complex nonlinear Schr\"odinger equation. Conversely, a Gross-Pitaevskii equation with complex on-site energies arises as the mean-field approximation of many-particle Lindblad dynamics rather than a complex extension of the Bose-Hubbard system. Here we address the question of how the mean-field description is modified in the presence of a complex-valued particle interaction term for a Bose-Hubbard dimer. We derive the mean-field equations of motion leading to nonlinear dissipative Bloch dynamics, related to a nontrivial complex generalisation of the nonlinear Schr\"odinger equation. The resulting dynamics are analysed in detail. It is shown that depending on the parameter values there can be up to six stationary states, and for small values of the interaction strength there are limit cycles. Furthermore, we show how a Gross-Pitaevskii equation with a complex interaction term can be derived as the mean-field approximation of a Bose-Hubbard dimer with an additional Lindblad term modelling two-particle losses. 
\end{abstract}

\section{Introduction}
The celebrated mean-field approximation arising in the context of multi-boson systems is closely related to the semiclassical approximation of single particle physics, where the particle number appears as an effective $\hbar^{-1}$. While many-particle systems are linear quantum systems on large tensor product Hilbert spaces, the mean-field approximation provides effective single particle descriptions on smaller Hilbert spaces. Interactions between the particles appear as nonlinearities in the mean-field description, leading to nonlinear Schr\"odinger equations of Gross-Pitaevskii type. The semiclassical relation between a full many-particle and an effective mean-field description has been intensively investigated in the literature. In particular, small systems, such as the Bose-Hubbard dimer, describing $N$ bosons in two modes with on-site interaction, are useful to gain deeper insights into this relation, as both many-particle and mean-field features are easily accessible. 

Recently there has been increasing interest in complexified quantum systems, arising as effective descriptions of lossy or scattering systems \cite{Mois11book}. Complex extensions of the Gross-Pitaevskii equation have also attracted considerable attention in the context of PT-symmetry, where a balanced distribution of loss and gain can lead to a quasi-closed behaviour of a system \cite{Muss08,08nhbh_s,10nhbh,Rame10,Li11,Zezy12,Cart12,Cart12b,12nhermBEC,Li12,Li13,Dast13}. A complex interaction strength (associated with a complex s-wave scattering length) in the Gross-Pitaevakii equation is often used to describe two-particle losses \cite{Mert07,Nick11}. However, how the mean-field approximation is modified in the presence of non-Hermiticities is a nontrivial question. It has been shown previously that the Gross-Pitaevskii equation with complex parameter values \textit{does not} arise as the mean-field approximation of a Bose-Hubbard system with complex parameter values, and \textit{vice versa}. Specifically, in \cite{08nhbh_s,10nhbh} it has been shown that the mean-field approximation for a Bose-Hubbard dimer with complex on-site energies is given by a different complexification of the nonlinear Schr\"odinger equation. Similarly it has been shown in \cite{Witt08} that the Gross-Pitaevskii equation with complex on-site energies can be viewed as the mean-field approximation of a many-particle dynamics governed by a Lindblad equation, modelling single-particle losses, where the imaginary parts of the on-site energies in the mean-field system is related to the strengths of particle losses from the two sites. Thus, a complex (non-Hermitian) generalisation of the many-particle system leads to a complex Gross-Pitaevskii equation, while a Lindblad dynamics of the many-particle system also leads to a (different) complex extension of the Gross-Pitaevskii equation, rather than a nonlinear generalisation of mixed state dynamics reminiscent to the many-particle Lindblad equation. Note that the mean-field limit of a Bose-Hubbard dimer with complex coupling constant has recently been studied in \cite{Zhon12}.

Here we address the question of how a complex value of the on-site interaction in the Bose-Hubbard dimer affects the mean-field approximation. We shall see that this modification leads to another nontrivial generalisation of the complex nonlinear Schr\"odinger equation,different from the previously considered cases of complex on-site energies or coupling constants.

The paper is organised as follows. We first briefly recall the mean-field approximation for the Hermitian Bose-Hubbard dimer in section \ref{sec_hermBH}. We then apply the approximation to the Bose-Hubbard dimer with complex interaction strength in section \ref{sec_MFapprox}. In section \ref{sec_MFdyn} we analyse the fixed-point structure and the dynamics of the resulting mean-field system in detail. We find analytic expressions for all fixed points and analyse their stability properties. To complement the investigation we discuss in section \ref{sec_LB} how the Gross-Pitaevskii equation with complex interaction parameter arises as the mean-field approximation of a many-particle Lindblad dynamics modelling two-particle losses. We finish with a summary in section \ref{sec_sum}. An appendix provides some details on calculations of expectation values of relevant operator products in condensed states.

\section{Mean-field and many-particle correspondence for the closed Bose-Hubbard dimer}
\label{sec_hermBH}
Cold atoms are \textit{a priori} many-particle quantum systems. For low temperatures and densities, however, when Bose-Einstein condensation occurs, they can be described by effective macroscopic single-particle wave functions. In the presence of particle interactions, typically a system is dynamically driven out of a condensed state. Nevertheless, the mean-field approximation, which assumes an initially condensed state to remain condensed for all times, can give valuable insights into the behaviour of the full many-particle system. While the many-particle description is a linear quantum system, the mean-field approximation induces effective nonlinearities due to the interaction between the particles. 

It is interesting to note that the mean-field approximation is formally closely related to the semiclassical limit of single-particle quantum mechanics. In this formulation the particle number appears as an effective inverse $\hbar$. The mean-field approximation is thus only strictly valid for initially condensed states in the singular limit of infinite particle numbers. Nevertheless, for finite particle numbers it is a good approximation for short times. Further, semiclassical techniques can be used to extract many-particle information from the mean-field approximation. The many-particle system shows typical ``quantum behaviour" on top of the ``classical" mean-field system, such as breakdowns and revivals in the dynamics. 

Let us now briefly review the relation between many-particle and mean-field description for Bose-Hubbard systems, specifically the Bose-Hubbard dimer. The Bose-Hubbard Hamiltonian 
\begin{eqnarray}\label{BH_Mmode}
 \hat{\mathcal{H}} & = & 
\sum_{j=1}^{M}\epsilon_j\hat{a}_j^{\dagger} \hat{a}_j+v\sum_{j=1}^{M-1} (\hat{a}_j^{\dagger} \hat{a}_{j+1} + \hat{a}_{j} \hat{a}_{j+1}^{\dagger}) + c\sum_{j=1}^{M}(\hat{a}_j^{\dagger} \hat{a}_j)^2 
\end{eqnarray}
can be used to model bosons in optical lattices with $M$ lattice sites. Here $\hat{a}_j$, $\hat{a}_j^{\dagger}$ are particle annihilation and creation operators for site $j$, and the parameters $v$,  $\epsilon_j$, and $c$ describe the coupling between adjacent sites, the on-site energy, and the on-site interaction, respectively. The total particle number operator $\hat{N} = \sum_{j=1}^M \hat{a}_j^{\dagger} \hat{a}_j$ commutes with the Hamiltonian, i.e. the dynamics conserves the total number of particles.

A generic state for $N$ particles in an $M$-mode Bose-Hubbard system is a vector in an $\frac{(M+N-1)!}{(M-1)! N!}$-dimensional Hilbert space. The subset of condensed states is given by symmetric product states where all particles populate the same single particle state, and is thus $M$-dimensional. The condensed states can be parametrised as 
\begin{equation}
\label{condensed_states_Mmode}
\ket{\psi_1, ..., \psi_M} = \frac1{\sqrt{N!}} \left(\sum_{j=1}^M \psi_j \hat{a}_j^{\dagger} \right)^N\!\!\! \ket{0, ..., 0},
 \end{equation}
where $\ket{0, ..., 0}$ denotes the vacuum state, i.e. the empty lattice. These states are in fact identical to the set of $SU(M)$ coherent states \cite{07phase,Zhan90}, and form a symplectic manifold. 

The mean-field approximation assumes that an initially condensed state stays condensed for all times. In the presence of interactions this is an approximation which is valid for finite time durations that scale with the particle number, similarly to the Ehrenfest time in semiclassical mechanics in dependence on $\hbar^{-1}$. 
In practical terms the approximation can be implemented in various ways. It can be viewed as constrained quantum dynamics \cite{Brod08b}, derived from a time-dependent variational principle \cite{Buon08} or directly from the dynamical equations for a set of observables by replacing operators with their expectation values in condensed states and taking the infinite particle number limit (while keeping the total interaction strength fixed). The coefficients $\psi_j$ are then interpreted as the components of the macroscopic mean-field wave function. The expectation value of the number operator in mode $j$ in the many-particle description is thus mapped to the population probability in the corresponding mode in the mean-field picture according to $\eval{\hat n_j}\to N|\psi_j|^2$. In this formulation the close relation to a semiclassical limit where quantum operators are replaced by classical observables in the limit of vanishing $\hbar$ is obvious \cite{Gnut98,08nhbh_s}.

In the Hermitian case the dynamics of many-particle observables are described by Heisenberg's equations of motion. The corresponding mean-field equations can be expressed directly in terms of canonical equations of motion where the classical Hamiltonian function is related to the expectation value of the many-particle Hamiltonian in condensed states $H=\langle \psi_j|\hat H|\psi_j\rangle$, and the quantum commutators are replaced by Poisson-brackets defined by the symplectic structure on the condensed state manifold \cite{Gnut98,Liu03}. 

For a two-mode system, such as the one considered in the present paper, the condensed states (\ref{condensed_states_Mmode}) reduce to 
\begin{equation}
\label{condensed_states}
\ket{\psi_1,\psi_2} = \frac1{\sqrt{N!}} ( \psi_1 \hat{a}_1^{\dagger} + \psi_2 \hat{a}_2^{\dagger} )^N \ket{0, 0}.
 \end{equation}
In analogy with the Bloch-representation of a single particle two-mode system, it is convenient to represent a many-particle two-mode system in terms of angular momentum operators $\hat{L}_x, \hat{L}_y, \hat{L}_z$ defined as \cite{Schw52}:
\begin{eqnarray}
\nonumber
\hat{L}_x &=  \frac1{2} (\hat{a}_1^{\dagger} \hat{a}_2 + \hat{a}_1 \hat{a}_2^{\dagger}), \\
\label{schwinger} 
\hat{L}_y &= \frac1{2i} (\hat{a}_1^{\dagger} \hat{a}_2 - \hat{a}_1 \hat{a}_2^{\dagger}), \\
\nonumber\hat{L}_z &= \frac1{2} (\hat{a}_1^{\dagger} \hat{a}_1 - \hat{a}_2^{\dagger}\hat{a}_2),
 \end{eqnarray}
obeying the usual $SU(2)$ commutation relations. The operator $\hat L_z$ represents the population imbalance of the two modes, and $\hat{L}_x$ and $\hat{L}_y$ measure their phase relation. 
  
In this formulation the equivalence between the set of condensed states (\ref{condensed_states}) and the  $SU(2)$ coherent states becomes obvious: $SU(2)$ coherent states are defined by 
\begin{equation}\label{su2cs}
\ket{\theta, \phi} = R(\theta, \phi) \ket{L},
\end{equation} 
where $R(\theta, \phi) = \exp[i\theta(\textbf{L}_x \sin\phi - \textbf{L}_y \cos\phi)]$, and $\ket{L}$ is the eigenstate of $\hat{L}_z$ with the highest quantum number. The equivalence of (\ref{condensed_states}) and (\ref{su2cs}) is achieved if we set
\begin{equation}\label{param}
\psi_1 = \sqrt{n} e^{- i\phi} \cos \frac{\theta}{2}, \qquad \psi_2 = \sqrt{n} \sin \frac{\theta}{2},
\end{equation}
where $n^N$ is the normalisation of the condensed many-particle state, which is usually chosen to be $1$. 

Let us now briefly demonstrate how the mean-field dynamics for the Bose-Hubbard dimer 
\begin{eqnarray}\label{BH_dimer_real}
 \hat{\mathcal{H}} & = & 
 \epsilon(\hat{a}_1^\dagger\hat a_1-\hat{a}_2^\dagger\hat a_2)+v(\hat{a}_1^{\dagger} \hat{a}_2 + \hat{a}_1 \hat{a}_2^{\dagger}) + c\left((\hat{a}_1^{\dagger} \hat{a}_1)^2 + (\hat{a}_2^{\dagger}\hat{a}_2)^2\right) \nonumber \\
&=& \epsilon(\hat{a}_1^\dagger\hat a_1-\hat{a}_2^\dagger\hat a_2)+v(\hat{a}_1^{\dagger} \hat{a}_2 + \hat{a}_1 \hat{a}_2^{\dagger}) + \frac{c}{2}(\hat{a}_1^{\dagger} \hat{a}_1 - \hat{a}_2^{\dagger}\hat{a}_2)^2+c\frac{\hat{N}^2}{2}
                                  \end{eqnarray}
can be obtained from the many-particle dynamics by replacing operators with their expectation values in coherent states. 

For this purpose we first express the Hamiltonian (\ref{BH_dimer_real}) in terms of the angular momentum operators as follows:
\begin{equation}\label{BH_dimer_real_L}
\hat{\mathcal{H}} = 2\epsilon L_z+ 2v\hat{L}_x + 2c\hat{L}_z^2+c\frac{\hat{N}^2}{2}.
\end{equation}
Note that this Hamiltonian is a special case of the Lipkin-Meshkov Glick model originally studied in the context of nuclear physics \cite{Lipk65,Mesh65,Glic65}.

The equations of motion for the angular momentum operators are given by
\begin{eqnarray}\label{eqtnsangmom_real}
\nonumber\frac{d}{dt}\eval{\hat{L}_x}  & = & -2\epsilon\eval{\hat L_y} - 2c \eval{[\hat{L}_y, \hat{L}_z]_+} \\
\frac{d}{dt}\eval{\hat{L}_y} & = & 2\epsilon\eval{\hat L_x}- 2v\eval{\hat{L}_z} + 2c \eval{[\hat{L}_x, \hat{L}_z]_+} \\
\nonumber\frac{d}{dt}\eval{\hat{L}_z} & = &  2v\eval{\hat{L}_y}.
\end{eqnarray}
We define the mean-field Bloch vector $\vec{s}$ by writing
\begin{equation} \label{bloch}
\eval{\hat{L}_i} = \frac{\bra{\theta,\phi} \hat{L}_i \ket{\theta,\phi}}{\braket{\theta,\phi}{\theta,\phi}} =:  Ns_i,
\end{equation}
with $i =x, y, z$. A short calculation then shows that
\begin{eqnarray}
s_x & = &\frac{1}{2}\sin\theta \cos\phi\nonumber \\
s_y  & = & \frac{1}{2} \sin\theta \sin\phi \\ 
s_z  & = & \frac{1}{2}\cos\theta,  \nonumber 
\end{eqnarray}
which defines the mean-field Bloch sphere of radius $\frac{1}{2}$.

Using the so-called D-algebra representation of the angular momentum operators \cite{Zhan90,Berm94,Gnut98} we can calculate the expectation values in $SU(2)$ coherent states of the anti-commutators appearing in (\ref{eqtnsangmom_real}) and find (see \ref{appendix_SU(2)} for details):
\begin{eqnarray} 
\label{expL1}\eval{[\hat{L}_i, \hat{L}_j]_+}= 2(1- \frac1{N})\eval{\hat{L}_i} \eval{\hat{L}_j} + \delta_{ij} \frac{N}{2},
\end{eqnarray}
for $ i = x, y, z$. 
Therefore, we find that the equations of motion for the Bloch vector in the condensed state approximation are given by
\begin{eqnarray}
\dot{s}_x                 & =& -2\epsilon s_y  - 4cN \left( 1 - \frac1{N} \right)s_y s_z   \nonumber   \\
\dot{s}_y   & =& 2\epsilon s_x+ 4cN \left( 1 - \frac1{N} \right)s_x s_z - 2vs_z \\
\dot{s}_z & =& 2v s_y\nonumber.
\end{eqnarray}
Taking the limit $N \rightarrow \infty$ while keeping $Nc = g$ fixed, we obtain the well-known mean-field equations of motion:
\begin{eqnarray}  
\dot{s}_x & = & -2\epsilon s_y - 4g s_y s_z \nonumber \\
\label{mfadyn}\dot{s}_y & = & 2\epsilon s_x + 4g s_x s_z - 2v s_z \\
\dot{s}_z & = & 2v s_y. \nonumber
\end{eqnarray}
These mean-field dynamics can be equivalently formulated in terms of a nonlinear Schr\"odinger equation of Gross-Pitaevskii type
\begin{eqnarray}\label{nlnhGPreal1}
i \dot\psi_1&=& (\epsilon+2g|\psi_1|^2)\psi_1+v\psi_2\\
\label{nlnhGPreal2}i\dot\psi_2&=& v \psi_1+ (-\epsilon+2g|\psi_2|^2)\psi_2.
\end{eqnarray}

Alternatively equations (\ref{nlnhGPreal1}) and (\ref{nlnhGPreal2}) can be obtained directly from the canonical structure inherited to the subspace of condensed states. This is achieved by evaluating the canonical equations of motion $i\dot\psi_j=\partial H/\partial \psi_j^*$, where $H$ is obtained from the many-particle Hamiltonian by replacing $\hat a_j\to\sqrt{N}\psi_j$ and $\hat a_j^\dagger\to\sqrt{N}\psi_j^*$, and dividing by the overall particle number:
\begin{equation}
H=\epsilon\left(|\psi_1|^2 + |\psi_2|^2\right)+v\left(\psi_1^* \psi_2 + \psi_1 \psi_2^*\right) + g\left(|\psi_1|^4 + |\psi_2|^4\right).
\end{equation}

The correspondence between the full many-particle description and the mean-field approximation for the Bose-Hubbard dimer, or equivalently the quantum-classical correspondence of the Meshkov-Lipkin-Glick model, has been the subject of many previous studies. It has been shown that the full many-particle dynamics of initially condensed states exhibits a rich structure of quantum phenomena such as breakdown and revivals on top of the mean-field dynamics (see, e.g., \cite{Milb97,Vard01a,Holt01a}). Further, semiclassical methods have been employed to reconstruct the many-particle spectrum from the mean-field dynamics \cite{07semiMP,Niss10,Simo12,Vida04b,Ribe07}.

In the case of complex parameter values the direct correspondence between the Bose-Hubbard dimer (\ref{BH_dimer_real}) and the discrete Gross-Pitaevskii equation (\ref{nlnhGPreal1}) and (\ref{nlnhGPreal2}) breaks down, as has been shown for the cases of complex on-site energies and coupling constants \cite{08nhbh_s,10nhbh,Witt08,Zhon12}. Let us now investigate how the mean-field description is modified in the presence of a complex interaction strength. 

\section{Mean-field approximation for the Bose-Hubbard dimer with complex interactions}
\label{sec_MFapprox}
For complex parameter values the mean-field approximation follows the same spirit as before, that is, an initially condensed state is assumed to stay condensed for all times. The only difference is that the many-particle Hamiltonian is no longer Hermitian. Therefore, the dynamical equations of motion for the many-particle observables are no longer given by Heisenberg's equations of motion \cite{Datt90b}. Instead, the dynamical equation for expectation values $\eval{\hat A}=\frac{\la\psi|\hat A|\psi\ra}{\la\psi|\psi\ra}$ is given by \cite{Datt90b,08nhbh_s}:
\begin{equation}\label{nonhermotion}
i\frac{d}{dt} \eval{\hat{A}} = \eval{[\hat{A}, \hat{H}]} - 2i \Delta^2_{\hat{A}\hat{\Gamma}},
\end{equation}
where $\Delta^2_{\hat{A}\hat{\Gamma}} := \eval{\frac1{2} [ \hat{A}, \hat{\Gamma}]_+} - \eval{\hat{A}}\eval{ \hat{\Gamma}}$ denotes the covariance, and $[\cdot, \cdot]$ and $[\cdot, \cdot]_+$ denote the commutator and the anti-commutator, respectively. Further, we assume that $\hat A$ does not explicitly depend on time. It has been  conjectured in \cite{10nhbh,10nhclass} that applying the condensed state approximation these equations lead to a generalised form of  canonical equations involving both the symplectic and the metric structure of the condensed state manifold. This is closely related to structures arising in the semiclassical limit of quantum dynamics generated by non-Hermitian Hamiltonians \cite{11nhcs}. The proof of this conjecture, however, goes beyond the scope of the present paper and will be the subject of a future study \cite{14nh_class_SUM}. 

Here we shall explicitly perform the mean-field approximation for the Bose-Hubbard dimer (\ref{BH_dimer_real}) with $c\to c-i\kappa$ by evaluating the many-particle equations of motion (\ref{nonhermotion}) for the angular momentum operators, replacing the expectation values with their values in $SU(2)$ coherent states (\ref{su2cs}) and perform the limit $N\to\infty$ while keeping the interaction strength fixed. 
For this purpose we need to express the Hamiltonian in terms of the angular momentum operators. In the following we shall focus on the case of a symmetric system with $\epsilon=0$. The Hamiltonian in terms of angular momentum operators is then given by
\begin{equation}\label{manypthamiltl}
\hat{\mathcal{H}} =  2v\hat{L}_x + 2c\hat{L}_z^2 -2i\kappa \hat{L}_z^2+\left(c-i\kappa\right)\frac{\hat{N}^2}{2},
\end{equation}
which is just (\ref{BH_dimer_real_L}) with the replacement $c\to c-i\kappa$.

From this we find the quantum equations of motion
\begin{eqnarray}\label{eqtnsangmom}
\nonumber\frac{d}{dt}\eval{\hat{L}_x}  & = &  - 2c \eval{[\hat{L}_y, \hat{L}_z]_+}  - 4\kappa \Delta^2_{\hat{L}_x\hat{L}_z^2} - \kappa \Delta^2_{\hat{L}_x\hat{N}^2}  \\
\frac{d}{dt}\eval{\hat{L}_y} & = & - 2v\eval{\hat{L}_z} + 2c \eval{[\hat{L}_x, \hat{L}_z]_+} - 4\kappa \Delta^2_{\hat{L}_y\hat{L}_z^2} - \kappa \Delta^2_{\hat{L}_y\hat{N}^2}  \\
\nonumber\frac{d}{dt}\eval{\hat{L}_z} & = &  2v\eval{\hat{L}_y}- 4\kappa \Delta^2_{\hat{L}_z\hat{L}_z^2} - \kappa \Delta^2_{\hat{L}_z\hat{N}^2} .
\end{eqnarray}
Further, the normalisation of the many-particle wave-function decays according to:
\begin{equation}
\label{MPnorm}
\frac{d}{dt}\braket{\Psi}{\Psi} = - 4 \kappa \left(\eval{\hat{L}_z^2}+\frac{\langle \hat N^2\rangle}{4}\right)\braket{\Psi}{\Psi}.
\end{equation}
The values of the additional covariances appearing in (\ref{eqtnsangmom}), as compared to (\ref{eqtnsangmom_real}), in $SU(2)$ coherent states are given by (see \ref{appendix_SU(2)} for details):
\begin{eqnarray} 
\Delta_{\hat{L}_i \hat{L}_z^2}^2 =-\frac{2}{N}\left( 1 - \frac1{N} \right) \eval{\hat L_i} \eval{\hat L_z}^2+ \delta_{iz}\frac{N}{2}\left( 1 - \frac1{N} \right) \eval{\hat L_z},\\
\label{expL3}\Delta_{\hat{L}_i \hat{N}^2}^2  = 0, 
\end{eqnarray}
for $ i = x, y, z$. 
Thus, we find that the equations of motion for the mean-field Bloch vector in the condensed state approximation are given by
\begin{eqnarray}
\dot{s}_x                 & =&  - 4cN \left( 1 - \frac1{N} \right)s_y s_z   + 8\kappa N\left( 1 - \frac1{N} \right)s_x s_z^2\nonumber   \\
\dot{s}_y   & =&   4cN \left( 1 - \frac1{N} \right)s_x s_z - 2vs_z+8\kappa N\left( 1 - \frac1{N} \right)s_y s_z^2 \\
\dot{s}_z & =& 2v s_y-2\kappa N \left[ \left( -4 + \frac4{N} \right)s_z^3  + \left( 1 - \frac{1}{N} \right) s_z \right]\nonumber.
\end{eqnarray}
Taking the limit $N \rightarrow \infty$ while keeping $Nc = g$ and $N\kappa = k$ fixed, we obtain the desired mean-field equations of motion:
\begin{eqnarray}  
\dot{s}_x & = &  - 4g s_y s_z +8k s_x s_z^2 \nonumber \\
\label{mfadyn}\dot{s}_y & = &   4g s_x s_z - 2v s_z + 8k s_y s_z^2 \\
\dot{s}_z & = & 2v s_y - 2k s_z (1- 4s_z^2). \nonumber
\end{eqnarray}
In addition the dynamics of the overall probability is given by:
\begin{equation} \label{MFAnorm}
 \dot{n} =  -4k\left(s_z^2 +\frac{1}{4}\right)n= -2\Gamma n,
\end{equation}
with $n^N = \braket{\Psi}{\Psi}$. 

We remark that these mean-field dynamics can equivalently be formulated in terms of a complex nonlinear Schr\"odinger equation of the form
\begin{eqnarray}\label{nlnhGP1}
i \dot\psi_1&=& \left(2(g\!-\!ik)\frac{|\psi_1|^2}{|\psi_1|^2+|\psi_2|^2}\!+\!ik\frac{|\psi_1|^4+|\psi_2|^4}{(|\psi_1|^2+|\psi_2|^2)^2}\right)\psi_1+v\psi_2\\
\label{nlnhGP2}i\dot\psi_2&=& v \psi_1+ \left(2(g\!-\!ik)\frac{|\psi_2|^2}{|\psi_1|^2+|\psi_2|^2} \!+\!ik\frac{|\psi_1|^4+|\psi_2|^4}{(|\psi_1|^2+|\psi_2|^2)^2}\right)\psi_2,
\end{eqnarray}
which differs considerably from a generalisation of (\ref{nlnhGPreal1}) and (\ref{nlnhGPreal2}) with complex $g$. Nevertheless, in the limit $k\to0$ the normalisation of the wave function is conserved, and the equations (\ref{nlnhGPreal1})-(\ref{nlnhGPreal2}) and (\ref{nlnhGP1})-(\ref{nlnhGP2}) become equivalent. 

We shall now proceed to analyse the resulting mean-field dynamics in detail.

\section{Mean-field dynamics}
\label{sec_MFdyn}
The equations of motion \rf{mfadyn} for the components of the Bloch vector $(s_x s_y, s_z)$ 
constitute a nonlinear, three-dimensional dynamical system. If the initial state is on the surface of the Bloch sphere, that is $ s_x^2 + s_y^2 + s_z^2 = \frac{1}{4}$, then it will stay on the sphere throughout the time evolution: $d/dt (s_x^2 + s_y^2 + s_z^2) = 0$. This can easily be seen in the following way:
\begin{equation}
\label{}
\frac{d}{dt} (s_x^2 + s_y^2 + s_z^2) = -16k s_z^2 \left( \frac1{4} - s_x^2 - s_y^2 - s_z^2 \right),
\end{equation}
which is equal to zero only if the Bloch vector is is initially on a sphere of radius $\frac{1}{2}$. In contrast, in the Hermitian case for which $k=0$, the Bloch vector remains confined to the surface of a sphere independently of the initial value of the radius.

The dynamics is organised according to the fixed points of the system, which correspond to the stationary solutions of the nonlinear Schr\"{o}dinger equation (\ref{nlnhGP1})-(\ref{nlnhGP2}). 
We shall show that depending on the parameter values there can be up to six stationary states, and provide analytic expressions for their positions. Furthermore, there can be limit cycles for sufficiently small values of the interaction strength. 

The fixed points are obtained by setting $\dot{s}_x$, $\dot{s}_y$ and $\dot{s}_z$ equal to zero in equations (\ref{mfadyn}). Setting $s_z = 0$, from the last equation in (\ref{mfadyn}) it follows that $s_y=0$, while $s_x$ is a free parameter. From the condition  $ s_x^2 + s_y^2 + s_z^2 = \frac{1}{4}$ we obtain the two solutions $(\pm 1/2, 0, 0)$. These two fixed points exist for arbitrary values of $k$ and $g$. Assuming that $s_z \neq 0$, on the other hand, we find:  
\begin{eqnarray}
\label{cacca1}s_x  =   \frac{g}{2v}(1 - 4s_z^2), \quad 
s_y  =   \frac{k}{v} s_z( 1 - 4s_z^2),
\end{eqnarray}
and 
\begin{eqnarray}
 - 2vs_z
+ \frac{2g^2}{v}( 1 - 4s_z^2)s_z + \frac{8 k^2}{v}( 1 - 4s_z^2) s_z^3 = 0, \label{robaccia}
\end{eqnarray}
which reduces to a biquadratic equation. Rescaling parameters such that $v = 1$ it takes the simple form
\begin{equation}\label{cacca3}
- g^2 + 1 + 4g^2 s_z^2 - 4k^2 s_z^2 + 16k^2 s_z^4 = 0.
\end{equation}
This yields the solutions:
\begin{equation}
\label{fixedptsonetwo}
 s_{1, \pm}  =  \left( \frac{g}{2}(1 - 4y_+), \mp k \sqrt{y_1}(1 - 4y_+), \pm \sqrt{y_+} \right),
 \end{equation}
and
\begin{equation}
 s_{2, \pm}  =  \left( \frac{g}{2}(1 - 4y_-), \mp k \sqrt{y_2}(1 - 4y_-), \pm \sqrt{y_-} \right),
\end{equation}
where
\begin{equation}
\label{fproot}
s_z^2 = y_{\pm} = \frac{k^2 - g^2 \pm \sqrt{g^4 + k^4 + 2g^2k^2 - 4k^2}}{8k^2}.
\end{equation}
However, these values only correspond to fixed points of our system if they are real and lie on the Bloch sphere of radius $\frac{1}{2}$.

Let us first consider for which values of $g$ and $k$, $y_{\pm}$ are real: This is the case when $P_k(g) := g^4 + k^4 + 2g^2k^2 - 4k^2 > 0$. The boundary of the corresponding region in parameter space is given by
\begin{equation}
\label{}
P_k(g) = 0 \Rightarrow g^2 + (k \mp 1)^2  = 1,
\end{equation}
which defines two circles of radius 1, centred at $(0, \pm 1)$ in the $(g, k)$-plane. Inside these two circular regions only the solutions $(\pm 1/2, 0, 0)$ will be allowed. Furthermore, $y_{\pm}$ needs to be positive for $s_z = \pm \sqrt{y_{\pm}}$ to be real. For $y_+$ this implies that 
\begin{equation}
\sqrt{g^4 + k^4 + 2g^2k^2 - 4k^2} > g^2 - k^2,
\end{equation}
which is always true when $|g| < |k|$, and holds only if $|g| > 1$ when $|g| > |k|$. On the other hand, we have that $y_- > 0$ if 
\begin{equation}
k^2 - g^2 > \sqrt{g^4 + k^4 + 2g^2k^2 - 4k^2},
\end{equation} 
which is satisfied only if $|g| < 1$. 

It can easily be verified that the fixed points $s_{1,2\pm}$ indeed fulfil the condition $s_x^2+s_y^2+s_z^2=\frac{1}{4}$ and thus lie on the Bloch sphere.

\begin{figure}[tb]\label{fig}
\begin{center}
\includegraphics[width=0.7\textwidth]{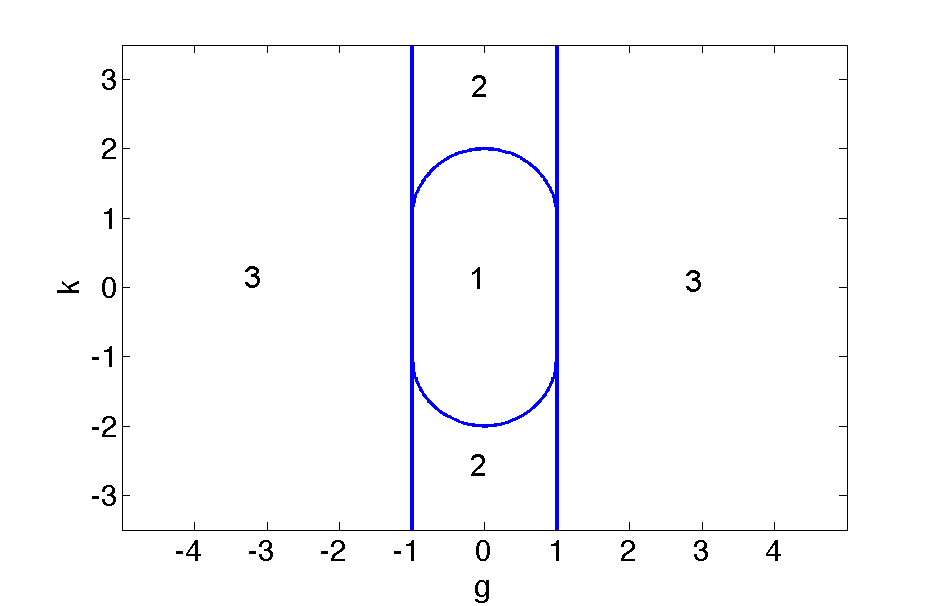}
\end{center}
\caption{\label{fig1} Regions of existence of the fixed points in the parameter plane $(g, k)$ for $v = 1$. There are two fixed points in region 1, six in region 2, and four in region 3.}
\end{figure}

The results obtained are summarised graphically in figure \ref{fig1}, representing the parameter space $(g, k)$ with three different regions of existence of the fixed points. Region 1 is where we have $P_k(g) < 0$, and hence the only two existing fixed points are $(\pm 1/2, 0, 0)$. Region 2 is defined by $P_k(g) > 0$ and $|g| < 1$, therefore we have all six fixed points $(\pm 1/2, 0, 0)$, $s_{1, \pm}$ and $s_{2, \pm}$. In region 3 we have $P_k(g) > 0$ and $|g| > 1$, therefore four fixed points exist, namely, $(\pm 1/2, 0, 0)$ and $s_{1, \pm}$. At the boundaries of these regions bifurcations take place.

Apart from the number of fixed points, the type of these fixed points characterises the resulting dynamics. Although (\ref{mfadyn}) is in general a three-dimensional system, here we are interested in the dynamics on the surface of the Bloch sphere, which is two-dimensional. 
Thus, there can be four main types of fixed points: \emph{sources}, from which the trajectories always depart; \emph{sinks}, towards which the trajectories converge; \emph{saddles}, from which all trajectories but one (the so-called \emph{stable manifold} of the saddle point) escape hyperbolically; and \emph{centres} or \emph{elliptic fixed points}, which are surrounded by closed, elliptical trajectories. The fixed points can be classified according to this scheme via the eigenvalues of the \emph{linearisation} or \emph{stability} matrix, defined by $J_{jk}=\frac{\partial \dot{s}_j}{\partial s_k}$, at each fixed point  \cite{Guck83,Arno88,Arno06,10nhbh}. 

Having set $v = 1$ in (\ref{mfadyn}) we obtain the following stability matrix:
\begin{equation}
  J =
  \left(\!\!\!\begin{array}{ccc} 
  8ks_z^2 & -4gs_z & - 4gs_y +16ks_x s_z \\
  4g s_z &  8k s_z^2 & 4gs_x - 2v + 16ks_y s_z \\
  -16ks_x s_z & 2v - 16ks_y s_z & -8k(s_x^2 + s_y^2)
  \end{array}\!\!\!\right),
 \end{equation}
where we have used that $s_x^2+s_y^2+s_z^2=\frac{1}{4}$.
From the behaviour of the eigenvalues of the stability matrix we find the following classification of fixed points for the upper half plane in the parameter space $g,k$  with $k>0$:
\begin{itemize}
\item Region 1: $(\pm 1/2, 0, 0)$ are both sinks. In addition there is an unstable limit cycle. For $g=0$ this coincides with the great circle with $s_x=0$. For $g\neq 0$ this orbit is slightly deformed.
\item Region 2: $(\pm 1/2, 0, 0)$ are sinks, $s_{1, \pm}$ are sources and $s_{2, \pm}$ are saddles.
\item Region 3: $(1/2, 0, 0)$ is a saddle, $( - 1/2, 0, 0)$ is a sink, and $s_{1, \pm}$ are sources.
\end{itemize}
Note that changing the sign of $k$, sinks and sources exchange and the unstable limit cycle in region 1 turns into a stable one: a change in the sign of $k$ converts attractors into repellers and vice-versa. On the division line between positive and negative values of $k$, that is, for $k=0$, all sinks and sources degenerate into elliptic fixed points and there is no longer a limit cycle. 

\begin{figure}[htb]\label{fig}
\begin{center}
\includegraphics[width=0.32\textwidth]{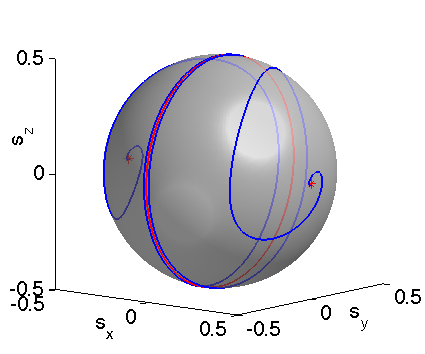}
\includegraphics[width=0.32\textwidth]{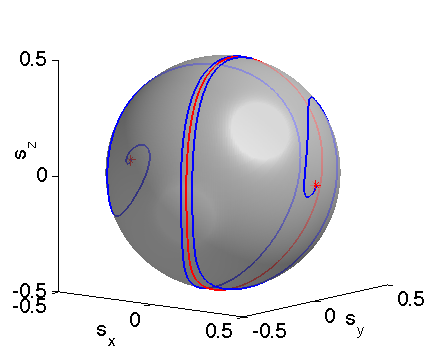}
\includegraphics[width=0.32\textwidth]{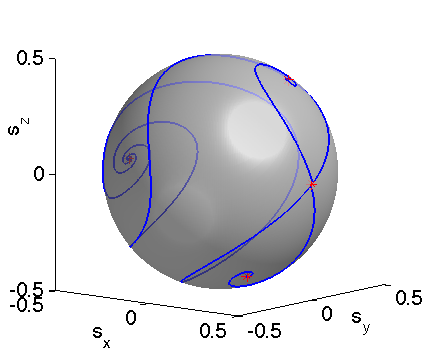}
\includegraphics[width=0.32\textwidth]{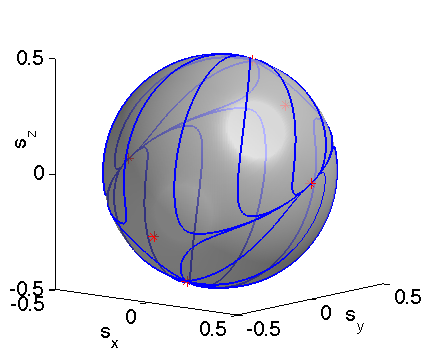}
\includegraphics[width=0.32\textwidth]{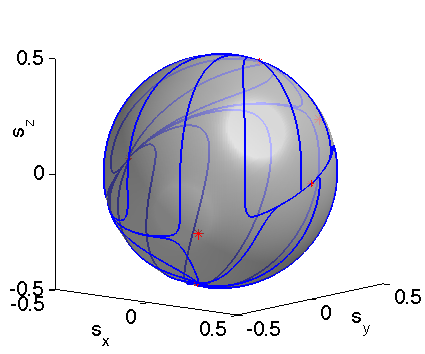}
\includegraphics[width=0.32\textwidth]{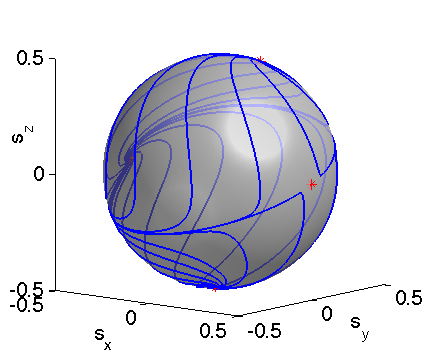}
\end{center}
\caption{\label{fig2} Dynamics on the Bloch sphere for $v=1$ and for different values of $k$ and $g$ (top panel: $k=1$ and different values of $g$, from left to right: $g=0$ (region 1), $g=0.5$ (region 1), and $g=1.5$ (region 3), bottom panel: $k=2.5$ and different values of $g$  from left to right: $g=0$ (region 2), $g=0.5$ (region 2), and $g=1.5$ (region 3)). Fixed points are marked by red stars, and limit cycles are plotted in red.}
\end{figure}

In figure \ref{fig2} examples of the dynamics on the Bloch sphere are shown for different parameter values in the three different regions. The first figure on the top left depicts the dynamics in region  1, for $g=0$. We observe the predicted limit cycle at $s_x=0$, which is unstable: For initial states with $s_x<0$ the system spirals into the fixed point at $(-\frac{1}{2},0,0)$; initial states with $s_x>0$ into the antipodal fixed point at $(\frac{1}{2},0,0)$. These two limiting states correspond to the symmetric and antisymmetric population of the two modes. The dynamics is mirror symmetric with respect to the limit cycle. For small values of $g$ we observe similar characteristics of the dynamics, however, the orbits are slightly deformed and the mirror symmetry is broken. An example is shown in the middle figure in the top panel. Increasing $g$ further above $g_{crit}=v$, two new fixed points emerge in a bifurcation reminiscent of the self-trapping bifurcation occuring for $k=0$: Depending on the sign of $g$ one of the fixed points at $s_z=0$ bifurcates into a saddle and two sources. The latter correspond to (unstable) stationary states with nonzero population difference. When the bifurcation occurs, the limit cycle vanishes simultaneously. Above the bifurcation the system eventually always approaches the remaining sink at $s_z=0$ irrespective of the initial position. An example of the resulting flow pattern is depicted in the  right figure in the top panel. 

Starting from region 1 and increasing the value of $k$, on the other hand, we observe a different scenario. At the boundary between region 1 and 2 the limit cycle vanishes and what remains are four fixed points along the ghost curve of the former limit cycle, two of which are sources, and two are saddle points. Above this transition, in region 2, the dynamics is governed by the flow from the two sources to the two sinks. That is, the system still asymptotically approaches one of the two fixed points at $s_z=0$, depending on the position of the initial state. However, instead of damped oscillations in the $s_z,s_y$ plane one now observes characteristically different orbits, as depicted for two examples in the left and middle figures in the bottom panel of figure 2. Approaching region 3 from region 2, the two saddle points approach one of the sinks at $s_z=0$, depending on the sign of $g$. At $g=v$ the two saddles meet at the sink and all three fixed points merge into a single saddle point. Above this transition the dynamics is dominated by a flow from the two sources to the one remaining sink at $s_z=0$, as depicted for an example in the lower right picture in figure 2. 

\begin{figure}[htb]\label{fig}
\begin{center}
\includegraphics[width=0.3\textwidth]{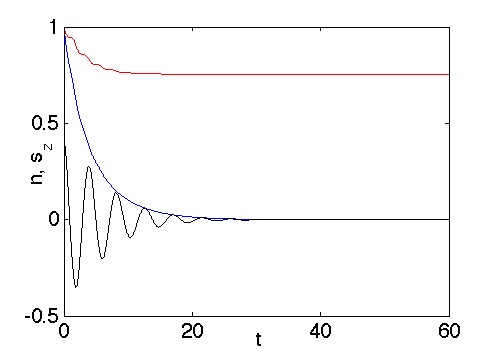}
\includegraphics[width=0.3\textwidth]{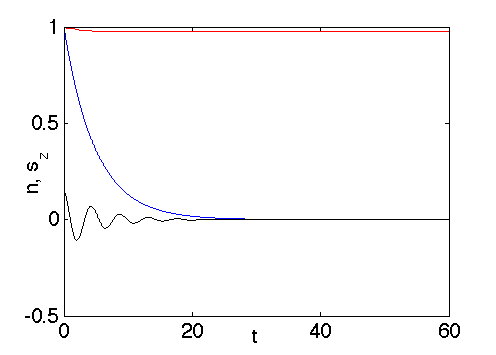}
\includegraphics[width=0.3\textwidth]{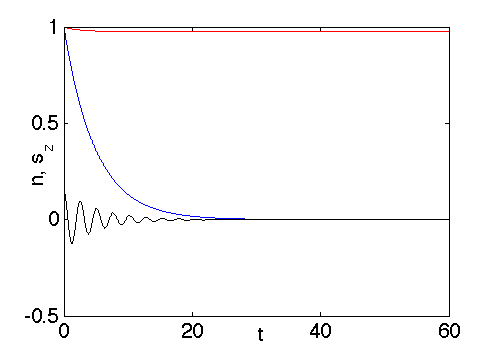}
\includegraphics[width=0.3\textwidth]{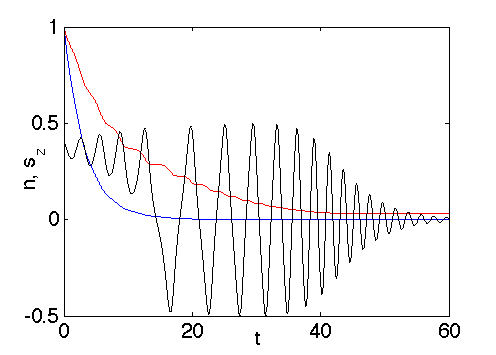}
\includegraphics[width=0.3\textwidth]{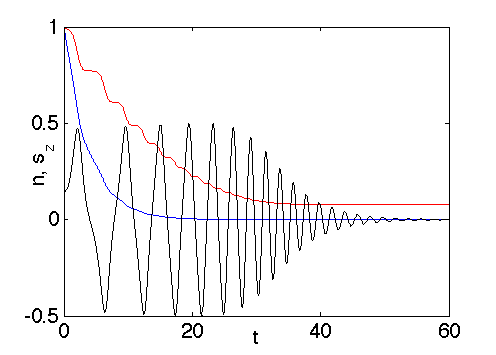}
\includegraphics[width=0.3\textwidth]{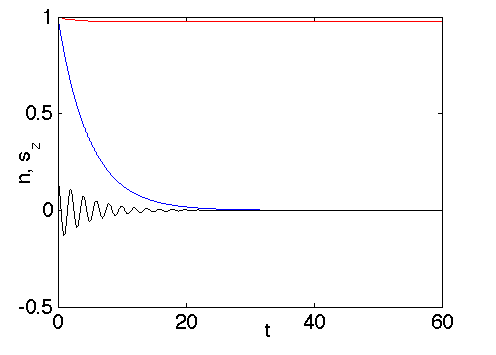}
\includegraphics[width=0.3\textwidth]{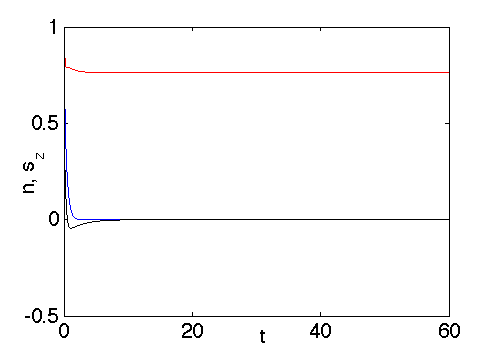}
\includegraphics[width=0.3\textwidth]{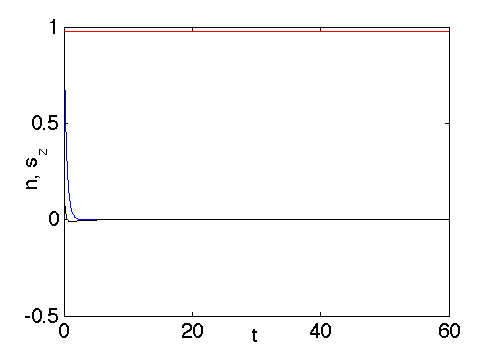}
\includegraphics[width=0.3\textwidth]{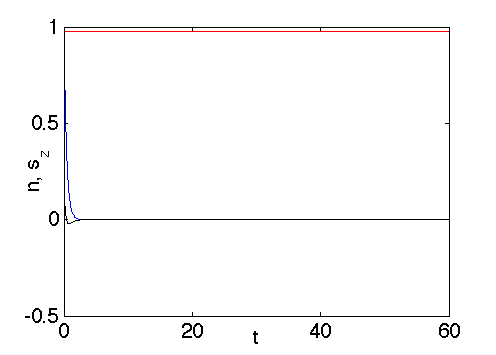}
\end{center}
\caption{\label{fig3} Dynamics of the population imbalance (black) and the norm (blue) as a function of time for $v=1$ for different initial values and different values of $k$ and $g$, corresponding to the three parameter regions in figure \ref{fig1}. The top panel corresponds to region 1 ($k=0.2$, $g=0.5$), the middle panel to region 3 ($k=0.2$, $g=1.5$), and the bottom panel to region 2 ($k=2.5$, $g=0.5$). The different columns correspond to different initial states: $(s_x,s_y,s_z)=(0.2939,0,0.4045)$ for the left column, $(s_x,s_y,s_z)=(0.4755,0,0,1545)$, for the middle and $(s_x,s_y,s_z)=(-0.4755,0,0,1545)$ for the right column. The red curve shows the norm without an overall exponential decay at rate $k$.}
\end{figure}

Let us finally comment on the resulting decay behaviour of the system. The decay of the norm (the survival probability), $\dot n=  -4k\left(s_z^2 +\frac{1}{4}\right)n$,
is locally exponential, the exponent varies with the square of the population imbalance. 
Thus, the different characteristic behaviours of the population imbalance dynamics in the different parameter regions are directly reflected in the decay behaviour. This is illustrated in figure \ref{fig3} where the decay of the norm is plotted in blue for three different initial states in each of the three parameter regions. For comparison the evolution of the population balance is depicted in the same plot in black. The different characteristics of the decay become more apparent when an overall exponential decay arising from the constant term $-i\frac{\kappa}{2}\hat{N}^2$ in the Hamiltonian is excluded, as shown by the red curves in the figure.  

In the limit of vanishing coupling between the two modes ($v=0$), we are left with a purely exponential decay. This might be somewhat surprising, as a complex interaction strength arises from a complex scattering lengths, which could be intuitively related to inelastic scattering and thus two-particle losses. However, while the decay arising here is clearly related to the two-particle interaction, it is considerably different from the decay typically associated with two-particle losses, leading to a non-exponential decay of the form $\dot n\propto -n^2$ \cite{Burt97,Li08}. The latter arises from a complex  interaction strength in the Gross-Pitaevskii equation (\ref{nlnhGPreal1})-(\ref{nlnhGPreal2}), which can be viewed as the mean-field limit of a many-particle Lindblad dynamics modelling two-particle losses, as we shall demonstrate in the following.  
 
\section{Nonlinear Schr\"odinger equation arising from Lindblad description of two-particle losses}
\label{sec_LB}
The main purpose of the present paper is to investigate how complex generalisations of the Bose-Hubbard Hamiltonian modify the corresponding mean-field approximation. To complement this investigation we shall now briefly discuss the opposite question of the many-particle model corresponding to a complex extension of the Gross-Pitaevskii equation, describing two-particle losses due to inelastic scattering. 

It should be evident that the physical effect of inelastic scattering in a many-particle system changes the expectation value of the particle number over time. A complex scattering length in the Bose-Hubbard Hamiltonian, on the other hand, leaves the expectation value of the particle number constant, while inducing a decay of the normalisation of the many-particle wave function. While the latter can be viewed as an effective measure of the number of remaining particles in the system, we have seen that the nature of the resulting decay qualitatively differs from the empirical effect of two-particle losses. A more physical way of modelling inelastic scattering in a Bose-Hubbard system makes use of a Lindblad description, where two-particle losses in the $j$-th mode are effectively described by the action of a Lindblad operator $a_j^2$ \cite{Li08,Syas08}:
\label{sec_MFLindblad}
\begin{equation}
\label{eqn_LB}
\dot{\hat\rho}=-i[\hat H,\rho]+\kappa\sum_{j=1,2} \left(\hat a_j^2\hat\rho \hat a_j^{\dagger 2} -\frac{1}{2}\hat a_j^{\dagger 2}\hat a_j^2\hat\rho-\frac{1}{2}\hat\rho \hat a_j^{\dagger 2}\hat a_j^2\right).
\end{equation}
Note that the imaginary part $\kappa$ of the scattering length appears as the loss coefficient in front of the Lindblad terms in this formulation. In this case the system does not generally stay in a pure state and is described by the density matrix $\hat\rho$. 
From (\ref{eqn_LB}) we find the following dynamics for the expectation values of angular momentum operators, defined in the usual way by $\eval{\hat L_j}=\Tr{\left(\hat L_j \hat \rho\right)}$:
\begin{eqnarray}
\nonumber\frac{d}{d t}\langle \hat L_x \rangle&=&  - 2c \eval{[\hat{L}_y, \hat{L}_z]_+}  -\kappa\left(\langle \hat N \hat L_x\rangle-\langle \hat L_x\rangle\right)\\
\frac{d}{d t}\langle \hat L_y \rangle&=&- 2v\eval{\hat{L}_z} + 2c \eval{[\hat{L}_x, \hat{L}_z]_+} -\kappa\left(\langle \hat N \hat L_y\rangle-\langle \hat L_y\rangle\right)\\
\nonumber \frac{d}{d t}\langle \hat L_z \rangle&=&2v\eval{\hat{L}_y}-2\kappa\left(\langle \hat N \hat L_z\rangle-\langle \hat L_z\rangle\right).
\end{eqnarray}
While the norm of the many-particle state stays conserved here, the expectation value of the total particle number acquires an explicit time dependence, governed by the dynamical equation
\begin{equation}
\frac{d}{d t}\langle \hat N \rangle=-2\kappa\left(\frac{1}{2}\langle \hat N^2\rangle+\langle \hat L_z^2\rangle-\langle \hat N\rangle\right).
\end{equation}

With these dynamical equations at hand, we can now perform the mean-field approximation as before, with the only difference that we have to allow for a varying particle number $N$ in the system. Following the procedure in \cite{Witt08}, effectively interpreting $N=\langle \hat N\rangle$ we find in the condensed state approximation:
\begin{eqnarray}
\nonumber\dot L_x &=&-4 c (1-\frac{1}{N})L_yL_z-\kappa\left(N -1\right)L_x,\\
\dot L_y&=&4 c (1-\frac{1}{N})L_yL_z-2vL_z-\kappa\left(N-1\right)L_y,\\
\nonumber\dot L_z&=&2vL_y-2\kappa\left(N-1\right)L_z,
\end{eqnarray}
where the $L_j$ are defined as the expectation values of the angular momentum operators in coherent states. For the dynamics of the total particle number we find  
\begin{equation}
\dot N=-2\kappa \left(1-\frac{1}{N}\right)\left( \frac{1}{2}N^2+2L_z^2\right).
\end{equation}
In analogy with (\ref{bloch}) we can define a mean-field Bloch vector $\vec{s}$, which is now normalised with respect to the initial particle number $N_0=\eval{\hat N(t=0)}$ as $N_0s_j:= L_j$, as well as the relative particle number $N_0n:=N$. Assuming that the particle number $N$ is initially large and stays large for all times of interest, we find the mean-field Bloch equations
\begin{eqnarray}
\label{Beqns_LB}
\nonumber \dot s_x &=&-4 g s_ys_z-k  ns_x\\
\dot s_y&=&4 gs_ys_z-2vs_z-kn s_y\\
\nonumber \dot s_z&=&2vs_y-2  k ns_z,
\end{eqnarray}
where $g:=N_0c$ and $k=N_0\kappa$. In contrast to the mean-field dynamics arising from the non-Hermitian Bose-Hubbard model, the Bloch vector is now constrained to a sphere of shrinking radius $\sum_j s_j^2=\frac{n^2}{4}$, with
\begin{equation} 
\label{eqn_norm_LB}
\dot n =-2k\left( \frac{1}{2}n^2+2s_z^2\right).
\end{equation}
Thus, the dynamics of the particle number shows the characteristic $\dot n\propto -n^2$ behaviour expected from two-particle losses.
The Bloch equations (\ref{Beqns_LB}) can be equivalently formulated in terms of the complex Gross-Pitaevskii equation
\begin{eqnarray}
\label{complex_GPE1}
i \dot\psi_1&=& (g-ik)|\psi_1|^2\psi_1+v\psi_2\\
\label{complex_GPE2}i\dot\psi_2&=& v \psi_1+ (g-ik)|\psi_2|^2\psi_2.
\end{eqnarray}
with the identification
\begin{eqnarray}
\nonumber
s_x &=  \frac1{2} (\psi_1^{*} \psi_2 + \psi_1 \psi_2^{*}), \\
s_y &= \frac1{2i}  (\psi_1^{*} \psi_2 + \psi_1 \psi_2^{*}), \\
\nonumber s_z &= \frac1{2} (\psi_1^{*} \psi_1 + \psi_2^* \psi_2),
 \end{eqnarray}
and $n=|\psi_1|^2+|\psi_2|^2$.
A further analysis of the dynamics generated by (\ref{complex_GPE1}) and (\ref{complex_GPE2}) would be an interesting topic for future investigations.  

\section{Summary}
\label{sec_sum}
We have analysed the mean-field approximation for a Bose-Hubbard dimer with complex interaction strength, and have shown that the resulting mean-field system differs from a simple complex extension of the Gross-Pitaevskii equation. The analysis of the mean-field dynamics showed that there can be up to six stationary states, which were explicitly calculated. We further analysed the stability of these stationary states, and the decay behaviour of the system. Finally we have shown how a Gross-Pitaevskii equation with complex interaction term can be derived as the result of a mean-field approximation of a Bose-Hubbard dimer with an additional Lindblad term modelling two-particle losses. 

\section*{Acknowledgments}
EMG gratefully acknowledges support via the Imperial College JRF scheme. The authors would like to thank Sven Gnutzmann and Dorje Brody for useful discussions.

\appendix
\section{Expectation values of angular momentum operator products in $SU(2)$ coherent states}
\label{appendix_SU(2)}

In the following it will be useful to express the $SU(2)$ coherent states in an alternative form using the complex parameter $\gamma : = e^{i \phi} \tan \frac{\theta}{2}$:
\begin{equation}\label{gammas}
\ket{\gamma} = (1 + \overline{\gamma}\gamma)^{- L} \exp (\gamma \hat{L}_-) \ket{L},
\end{equation}
where $\hat{L}_- = \hat{L}_x - i \hat{L}_y$.

The equivalence of (\ref{gammas}) and (\ref{su2cs}) can be verified by applying the Baker-Campbell-Hausdorff formula to rewrite \cite{Zhan90}:
\begin{equation}\label{gammas1}
\hat{R} = \exp ( \gamma \hat{L}_- ) \exp [- \ln (1 + \overline{\gamma}\gamma)\hat{L}_z] \exp ( -\overline{\gamma}\hat{L}_+),
\end{equation}
and using that $\exp ( -\overline{\gamma}\hat{L}_+) \ket{L} = \ket{L}$, and $\exp [- \ln (1 + \overline{\gamma}\gamma)\hat{L}_z] \ket{L} = (1 + \overline{\gamma}\gamma)^{- L} \ket{L}$.

Furthermore, it will be useful to distinguish a non-normalised version of the $SU(2)$ coherent states defined as:
\begin{equation}\label{gammas2}
|\ket{\gamma} = \exp (\gamma \hat{L}_-) \ket{L},
\end{equation}
such that $\ket{\gamma} = (1 + \overline{\gamma}\gamma)^{- L} |\ket{\gamma}$.
The normalisation $K$ is then given by
\begin{equation} 
K(\gamma, \bar{\gamma}) := \dbraket{\gamma}{\gamma} = (1 + |\gamma|)^N.
\end{equation}

The action of an operator $\hat{L}_i$ on a state $\dket{\gamma}$ can be described by means of a differential operator $\hat{d}_i$ defined in terms of $\gamma$ (for details see, e.g., \cite{Gnut98} and \cite{Berm94}):
\begin{eqnarray} 
\dbra{\gamma} \hat{L}_+ \dket{\gamma} & = & \dbra{\gamma} \hat{d}_+ \dket{\gamma} =  \hat{d}_+ \dbraket{\gamma}{\gamma} =  \left( \gamma N - \gamma^2 \frac{\partial}{\partial \gamma} \right) \dbraket{\gamma}{\gamma}\nonumber  \\
                                                                         & := & (\hat{d}_+^0 + \hat{d}_+^D) \dbraket{\gamma}{\gamma} \\
\dbra{\gamma} \hat{L}_- \dket{\gamma} & = & \dbra{\gamma} \hat{d}_- \dket{\gamma} = \hat{d}_- \dbraket{\gamma}{\gamma} = \frac{\partial}{\partial \gamma} \dbraket{\gamma}{\gamma} \nonumber \\
                                                                      & := & (\hat{d}_-^0 + \hat{d}_-^D) \dbraket{\gamma}{\gamma} \\
\dbra{\gamma} \hat{L}_z \dket{\gamma} & = & \dbra{\gamma} \hat{d}_z \dket{\gamma} = \hat{d}_z \dbraket{\gamma}{\gamma} = \left( \frac{N}{2} - \gamma \frac{\partial}{\partial \gamma} \right) \dbraket{\gamma}{\gamma} \nonumber \\
                                                                      & := & (\hat{d}_z^0 + \hat{d}_z^D) \dbraket{\gamma}{\gamma}.
\end{eqnarray}
Thus, one verifies that
\begin{eqnarray} 
 \eval{\hat{L}_+} & = & \frac{1}{K} \hat{d}_+ K = \frac{N \gamma}{1 + |\gamma|^2}\\
\eval{\hat{L}_-} & = & \frac{1}{K} \hat{d}_- K = \frac{N \bar{\gamma}}{1 + |\gamma|^2}\\
s_z & = & \frac{1}{N}\frac{1}{K} \hat{d}_z K = \frac{1}{2} \left( \frac{1 - |\gamma|^2 }{1 +|\gamma|^2} \right),
\end{eqnarray}
and therefore
\begin{eqnarray} 
s_x & = & \frac{1}{2} \frac{1}{N}\left( \frac{1}{K} \hat{d}_+ K + \frac{1}{K} \hat{d}_-  K \right) =  \frac{1}{2} \left( \frac{\gamma + \bar{\gamma}}{1 + |\gamma|^2} \right) \\
s_y & = &  \frac{1}{2i} \frac{1}{N} \left( \frac{1}{K} \hat{d}_+ K - \frac{1}{K} \hat{d}_- K \right) =  \frac{1}{2i} \left( \frac{\gamma - \bar{\gamma}}{1 + |\gamma|^2} \right).
\end{eqnarray}
With $\gamma : = e^{i \phi} \tan \frac{\theta}{2}$ this reduces to the spherical coordinates representation of the Bloch sphere.

With the aim to evaluate terms arising from the anticommutators in (\ref{eqtnsangmom}), we make use of the relations:
\begin{eqnarray} \label{star}
\bra{\gamma} \hat{f} \hat{L}_j \ket{\gamma} & = & \overline{\bra{\gamma} \hat{L}_j \hat{f} \ket{\gamma}}   =  \frac{\dbra{\gamma} \hat{f} \hat{L}_j \dket{\gamma}}{\dbraket{\gamma}{\gamma}} \nonumber \\
                                                                       & = & \frac{1}{K} \hat{d}_j K \frac{\dbra{\gamma} \hat{f} \dket{\gamma}}{\dbraket{\gamma}{\gamma}} \nonumber \\
                                                                       & = & \frac{1}{K} ( \hat{d}_j^0 + \hat{d}_j^D) (K f) \nonumber\\
                                                                        & = & Ns_j f + \hat{d}_j^D f, 
\end{eqnarray}
where $\hat{f}$ is an arbitrary Taylor expandable function of the angular momentum operators and $f := \frac{\dbra{\gamma} \hat{f} \dket{\gamma}}{\dbraket{\gamma}{\gamma}} = \eval{\hat{f}}$.

From the above it immediately follows 
\begin{equation} 
\eval{[\hat{L}_i, \hat{L}_j]_+} = 2(1- \frac1{N})\eval{\hat{L}_i} \eval{\hat{L}_j} + \delta_{ij} \frac{N}{2},
\end{equation}
\begin{equation} 
\eval{[\hat{L}_i, \hat{N}]_+} = 2\eval{\hat N}\eval{\hat{L}_i}.
\end{equation}
and
\begin{equation} 
\eval{[\hat{L}_i, \hat{N}^2]_+} = 2\eval{\hat N^2}\eval{\hat{L}_i}= 2\eval{\hat N}^2\eval{\hat{L}_i}.
\end{equation}
Further, we find
\begin{equation} 
 \eval{\hat{L}_z^3}  =  N s_z \eval{\hat{L}_z^2} + \hat{d}_z^D \eval{\hat{L}_z^2}  =  N^3\left( 1 - \frac3{N} + \frac2{N^2} \right) s_z^3 + N^2 \left( \frac3{4} - \frac1{2N} \right) s_z,
\end{equation}
\begin{eqnarray} 
\eval{\hat{L}_z^2  \hat{L}_x} & = & \overline{\eval{\hat{L}_x \hat{L}_z^2}} \nonumber \\
                              & = & N s_x \eval{\hat{L}_z^2} + \hat{d}_x^D \eval{\hat{L}_z^2} \nonumber \\
                              & = & N ^3  \left( 1 - \frac1{N} \right) s_x s_z^2 + \frac{N^2}{4} s_x + N^2  \left( 1 - \frac1{N} \right) \hat{d}_x^D s_z^2 \nonumber \\
                              & = & N ^3  \left( 1 - \frac1{N} \right) s_x s_z^2 + \frac{N^2}{4} s_x + N^2  \left( 1 - \frac1{N} \right) (- 2 s_x s_z^2 + i s_y s_z) \nonumber \\
                              & = & N^3 \left( 1 - \frac3{N} + \frac1{N^2} \right) s_x s_z^2 + \frac{N^2}{4} s_x + iN^2 \left( 1 - \frac1{N} \right) s_y s_z,
\end{eqnarray}
and
\begin{eqnarray} 
\eval{\hat{L}_z^2  \hat{L}_y} & = & \overline{\eval{\hat{L}_y \hat{L}_z^2}} \nonumber\\
                               & = & N s_y \eval{\hat{L}_z^2} + \hat{d}_y^D \eval{\hat{L}_z^2} \nonumber \\
                              & = & N ^3  \left( 1 - \frac1{N} \right) s_y s_z^2 + \frac{N^2}{4} s_y + N^2  \left( 1 - \frac1{N} \right) \hat{d}_y^D s_z^2 \nonumber \\
                              & = & N ^3  \left( 1 - \frac1{N} \right) s_y s_z^2 + \frac{N^2}{4} s_y + N^2  \left( 1 - \frac1{N} \right) (\frac1{i} s_x s_z - 2 s_y s_z^2) \nonumber \\
                              & = & N^3 \left( 1 - \frac3{N} + \frac1{N^2} \right) s_y s_z^2 + \frac{N^2}{4} s_y + \frac{N}{i}^2 \left( 1 - \frac1{N} \right) s_x s_.
\end{eqnarray}

From this we can now derive the following terms:
\begin{eqnarray} 
\Delta_{\hat{L}_i \hat{N}^2}^2 & = & \eval{\frac1{2} [\hat{L}_i, \hat{N}^2]_+} - \eval{\hat{L}_i}\eval{\hat{N}^2} \nonumber \\
                               & = &  \eval{\hat{L}_i \hat{N}^2} - \eval{\hat{L}_i}\eval{\hat{N}^2} \nonumber \\
                               & = &  \eval{\hat{L}_i}\eval{\hat{N}^2} - \eval{\hat{L}_i}\eval{\hat N^2} = 0, \quad \mbox{for} \quad i = x, y, z,
\end{eqnarray}
\begin{eqnarray} 
\Delta_{\hat{L}_x \hat{L}_z^2}^2 & = & \eval{\frac{1}{2} [\hat{L}_x, \hat{L}_z^2]_+} - \eval{\hat{L}_x}\eval{\hat{L}_z^2}  =  \Re \left( \eval{\hat{L}_z^2 \hat{L}_x} \right) - \eval{\hat{L}_x}\eval{\hat{L}_z^2} \nonumber \\
                                     & = &  N^3 \left( 1 - \frac3{N} + \frac2{N^2} \right) s_x s_z^2 + \frac{N^2}{4} s_x - N^3\left( 1 - \frac1{N} \right)s_x s_z^2 - \frac{N^2}{4} s_x \nonumber \\
                                    & = & N^3\left( - \frac2{N} + \frac2{N^2} \right) s_x s_z^2,
\end{eqnarray}
\begin{eqnarray}
\Delta_{\hat{L}_y \hat{L}_z^2}^2 & = & \eval{\frac{1}{2} [\hat{L}_y, \hat{L}_z^2]_+} - \eval{\hat{L}_y}\eval{\hat{L}_z^2}  =  \Re \left( \eval{\hat{L}_z^2 \hat{L}_y} \right) - \eval{\hat{L}_y}\eval{\hat{L}_z^2} \nonumber \\
                                     & = &  N^3 \left( 1 - \frac3{N} + \frac2{N^2} \right) s_y s_z^2 + \frac{N^2}{4} s_y - N^3\left( 1 - \frac1{N} \right)s_y s_z^2 - \frac{N^2}{4} s_y \nonumber \\
                                    & = & N^3\left( - \frac2{N} + \frac2{N^2} \right) s_y s_z^2,
\end{eqnarray}
and
\begin{eqnarray}
\Delta_{\hat{L}_z \hat{L}_z^2}^2 & = & \eval{\hat{L}_z^3} - \eval{\hat{L}_z}\eval{\hat{L}_z^2} \nonumber \\
                                     & = & N^3 \left( - \frac2{N} + \frac2{N^2} \right) s_z^3 + \frac{N^2}{2}\left( 1 - \frac1{N} \right) s_z.
\end{eqnarray}

Similarly products of higher powers of the angular momentum operators can be iteratively related to the mean-field observables $s_j$.

\end{document}